\documentclass[12pt,preprint]{aastex}
\begin{document}
\def\eg{{\it e.g.}}
\def\ie{{\it i.e.}}
\newcommand{\tnm}[1]{\tablenotemark{#1}}
\newbox\grsign
\setbox\grsign=\hbox{$>$}
\newdimen\grdimen
\grdimen=\ht\grsign
\newbox\simlessbox
\newbox\simgreatbox
\setbox\simgreatbox=\hbox{\raise.5ex\hbox{$>$}\llap
     {\lower.5ex\hbox{$\sim$}}}\ht1=\grdimen\dp1=0pt
\setbox\simlessbox=\hbox{\raise.5ex\hbox{$<$}\llap
     {\lower.5ex\hbox{$\sim$}}}\ht2=\grdimen\dp2=0pt
\def\simgreat{\mathrel{\copy\simgreatbox}}
\def\simless{\mathrel{\copy\simlessbox}}

\title{Spectroscopic Survey of Red Giants in the SMC. I:  Kinematics
\altaffilmark{1}}

\author{Jason Harris \& Dennis Zaritsky}
\affil{Steward Observatory}
\affil{933 North Cherry Ave., Tucson, AZ, 85721}
\email{jharris@as.arizona.edu, dzaritsky@as.arizona.edu}

\altaffiltext{1}{This paper includes data gathered with the 6.5 meter
  Magellan Telescopes located at Las Campanas Observatory, Chile.}

\begin{abstract}
  We present a spectroscopic survey of 2046 red giant stars,
  distributed over the central 4~kpc$\times$2~kpc of the Small
  Magellanic Cloud (SMC). After fitting and removing a small velocity
  gradient across the SMC ($7.9$~km~s$^{-1}$~deg$^{-1}$ oriented at
  10$^\circ$ E of N), we measure an rms velocity scatter of
  $27.5\pm0.5$~km~s$^{-1}$. The line of sight velocity distribution is
  well-characterized by a Gaussian and the velocity dispersion profile
  is nearly constant as a function of radius. We find no kinematic
  evidence of tidal disturbances.  Without a high-precision
  measurement of the SMC's proper motion, it is not possible to
  constrain the SMC's true rotation speed from our measured
  radial-velocity gradient.  However, even with conservative
  assumptions, we find that $v < \sigma$ and hence that the SMC is
  primarily supported by its velocity dispersion. We find that the
  shape of the SMC, as measured from the analysis of the spatial
  distribution of its red giant stars, is consistent with the degree
  of rotational flattening expected for the range of allowed
  $v/\sigma$ values. As such, the properties of the SMC are consistent
  with similar low luminosity spheroidal systems.  We conclude that
  the SMC is primarily a low luminosity spheroid whose irregular
  visual appearance is dominated by recent star formation. A simple
  virial analysis using the measured kinematics implies an enclosed
  mass within 1.6~kpc of between 1.4 and $1.9\times10^9 M_\odot$, and
  a less well constrained mass within 3~kpc of between 2.7 and
  $5.1\times10^9 M_\odot$.
\end{abstract}


\section{Introduction}\label{sec:intro}

There is an abundant and growing body of evidence that gravitational 
interactions have been critical in the evolution of the Magellanic 
Clouds. One prominent piece of evidence is the existence of the 
Magellanic Stream \citep{mat74}, a roughly contiguous filament of 
neutral hydrogen that stretches more than 150~kpc behind the Clouds 
and which numerical simulations suggest was stripped by a recent 
interaction among the Clouds and the Milky Way \citep{gn96}.  The 
Clouds are the nearest  galaxies that are globally affected by 
interactions yet are far from being disrupted. As such, they are 
detailed examples of galaxies expected to be common in a hierarchical 
model of structure formation and may provide general insights on how 
such weak interactions affect both the observed characteristics and 
evolution of galaxies.

For the Small Magellanic Cloud (SMC), we have presented results on 
the effects of the interactions on both its global properties and its 
evolution. First, we demonstrated that the irregular appearance of
the SMC is due solely to recent star formation, because its older 
stellar populations are distributed in a regular, smooth ellipsoid 
\citep{zar00}. This result cautions us that galaxies that appear to 
be irregular, particularly low mass galaxies like the SMC, may indeed 
have quite a different underlying structure --- a conclusion that is 
obviously critical to such considerations as whether dIrr's can 
evolve into dE's.  Second, we found that the SMC has experienced 
periodic episodes of global enhanced star formation, suggesting 
triggering by perigalactic passes with the Large Magellanic Cloud 
and with the Milky Way \citep{hz04,zh04}.  These results demonstrate 
for one galaxy, and suggest for others, that basic observational 
parameters such as morphology, luminosity, and color are 
significantly affected by star formation induced by weak 
interactions.  Here, we investigate (1) whether such interactions 
also induce detectable kinematic disturbances, and so whether common 
kinematical analyses, such as the classification of galaxies by the 
ratio of rotational velocity to velocity dispersion and the 
application of the virial theorem to estimate the mass, are valid for 
galaxies affected by weak interactions and (2) whether the 
classification of the SMC as a dE based on the distribution of the 
older stars \citep{zar00} is substantiated by the kinematics.

The conventional picture of the SMC as a disturbed galaxy is derived 
from the kinematics of stars in its outer regions 
\citep{hat93,hat97}, and from a large observed line-of-sight depth 
\citep{hh89, gh91, cro01}.  It is important to note that these 
line-of-sight depth studies also generally sample the outer regions 
of the SMC.  The depth of the central regions of the SMC has been 
probed using Cepheid variables, but there seems to be no consensus 
on the resulting depth.  While some authors find a large 
line-of-sight depth of around 20~kpc \citep{mat86, mat88, gro00}, 
others conclude the depth is only 6--8~kpc \citep{wel87, lah05}.  
Much of the controversy can be resolved by simply recognizing that 
the term ``depth'' is used differently by the two groups of authors.  
Those who define the depth as the full front-to-back extent of the 
tracer population obtain a depth around 20~kpc, while those who 
define the depth as the full width at half-maximum of the 
distribution of distances obtain a much smaller depth around 
6--8~kpc.  The conclusion to be drawn from this body of work is that 
while the outer regions of the SMC may be extended along the line of 
sight (up to 20~kpc), especially toward the Northeast, its central 
region seems to be only modestly extended, with a depth of 6--8~kpc.

Our knowledge of the internal kinematics of the main body of the SMC, 
the region that would typically be observed in more distant galaxies, 
is surprisingly limited, considering the SMC's proximity and
suspected complex internal structure.  Previous studies have 
determined mean radial velocities and velocity dispersions, based on 
modest numbers of tracer populations:  44 central planetary nebulae 
\citep{dop85}, 12 outer RGB stars \citep{sun86}, 131 central Carbon 
stars \citep{har89}, 30 outer red clump stars \citep{hat93}, 
and 70 outer Carbon stars \citep{hat97}. In each case, the observed 
sample was too small to support analysis beyond the mean velocity and 
velocity dispersion of the sample.

In this paper, we present a spectroscopic survey of over 2000 red 
giant branch stars in the main body of the SMC, using the Inamori 
Magellan Areal Camera and Spectrograph \citep[IMACS;\ ][]{big98} at 
the Magellan Baade Telescope at Las Campanas Observatory in Chile.  
In Section~\ref{sec:obsdata}, we describe the observations and data 
reduction.  In Section~\ref{sec:kinematics}, we discuss whether the 
SMC kinematics follow the expectations for a dE and whether we find 
any kinematic signature of the weak interactions the SMC has recently 
experienced.  We summarize the results in Section~\ref{sec:summary}.

In addition to obtaining radial-velocity kinematics of a large 
sample of SMC field stars, the observations were also designed to 
yield metallicities based on the equivalent width of the 
Ca~{\small II} triplet feature.  The metallicities of our sample and 
their constraint on the chemical enrichment history of the SMC are 
presented in a subsequent paper \citep{har05}.

\section{Observations and Data Reduction}\label{sec:obsdata}

\subsection{Sample Selection}\label{sec:sample}

We select candidate red giant branch (RGB) stars photometrically from
our Magellanic Clouds Photometric Survey (MCPS) catalog, which
contains subarcsecond astrometry and $UBVI$ photometry for over 6
million SMC stars \citep{zar02}. We employ the following photometric 
criteria to select candidate RGB stars: 
$$15.4 < I < 16.4$$ 
and 
$$(20.50 -4.0\times(V-I)) < I < (22.0 - 4.0\times(V-I))$$ 

\begin{figure}[h]
\plotone{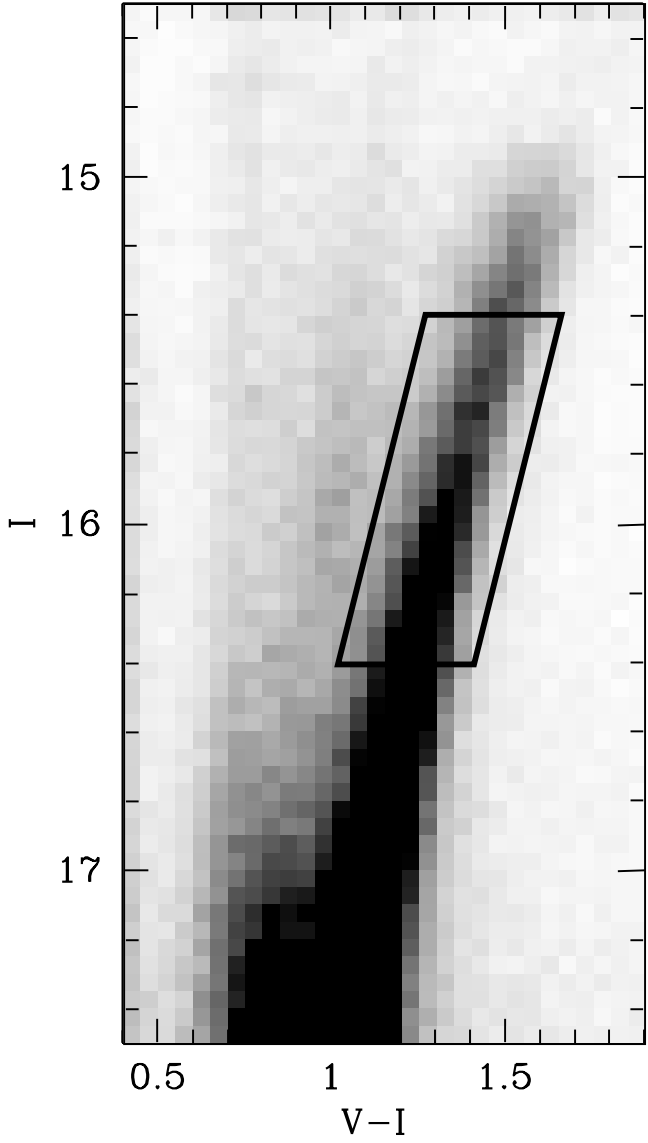}
\caption{ A V-I, I Hess diagram showing MCPS photometry of the red
giant branch in the Small Magellanic Cloud. The black box illustrates
the photometric criteria used to select 30,000 bright red giant
candidates for our spectroscopic sample.  The criteria were chosen to
include the full range of RGB colors, and to avoid the tip of the RGB,
where deep mixing may alter surface abundances. \label{fig:cmd} }
\end{figure}

\noindent These photometric criteria are plotted on a $V-I$,$I$ Hess 
diagram of the SMC in Figure~\ref{fig:cmd}. The bright limit 
($I>15.4$~mag) excludes stars near the tip of the RGB, where deep 
mixing may alter the surface abundances.  The faint limit 
($I<16.4$~mag) ensures that we can obtain spectra with 
S/N$\simgreat20$ in a 40-minute exposure. The diagonal cuts were 
designed to be roughly parallel to the fiducial RGB of the SMC 
stellar population.  Approximately 34,000 candidate RGB stars 
throughout the $4.5^\circ \times 4^\circ$ region covered by the MCPS 
satisfy these criteria.

In addition to RGB stars, we also select a sample of alignment stars 
for precision positioning of the slitmasks.  A key consideration for 
slitmask alignment is that the alignment stars are drawn from the 
same catalog as our target stars and hence share the same astrometric 
system.  Because our SMC fields are much more crowded than a typical 
pointing on the sky,  we restrict our alignment-star selection to 
a narrow range of brightnesses ($17.0 < I < 17.5$), at the bright end 
of the range of magnitudes recommended in the IMACS user manual.  We 
also impose a color cut ($0.5 < (V-I) < 1.5$), to further reduce 
the large number of candidates.  Lastly, we reject crowded objects, 
those that are less than 20\arcsec\ from neighboring stars brighter 
than $I=17.5$~mag.  We are left with approximately 13000 uncrowded 
alignment star candidates throughout the SMC. 

\begin{figure}[h]
\plotone{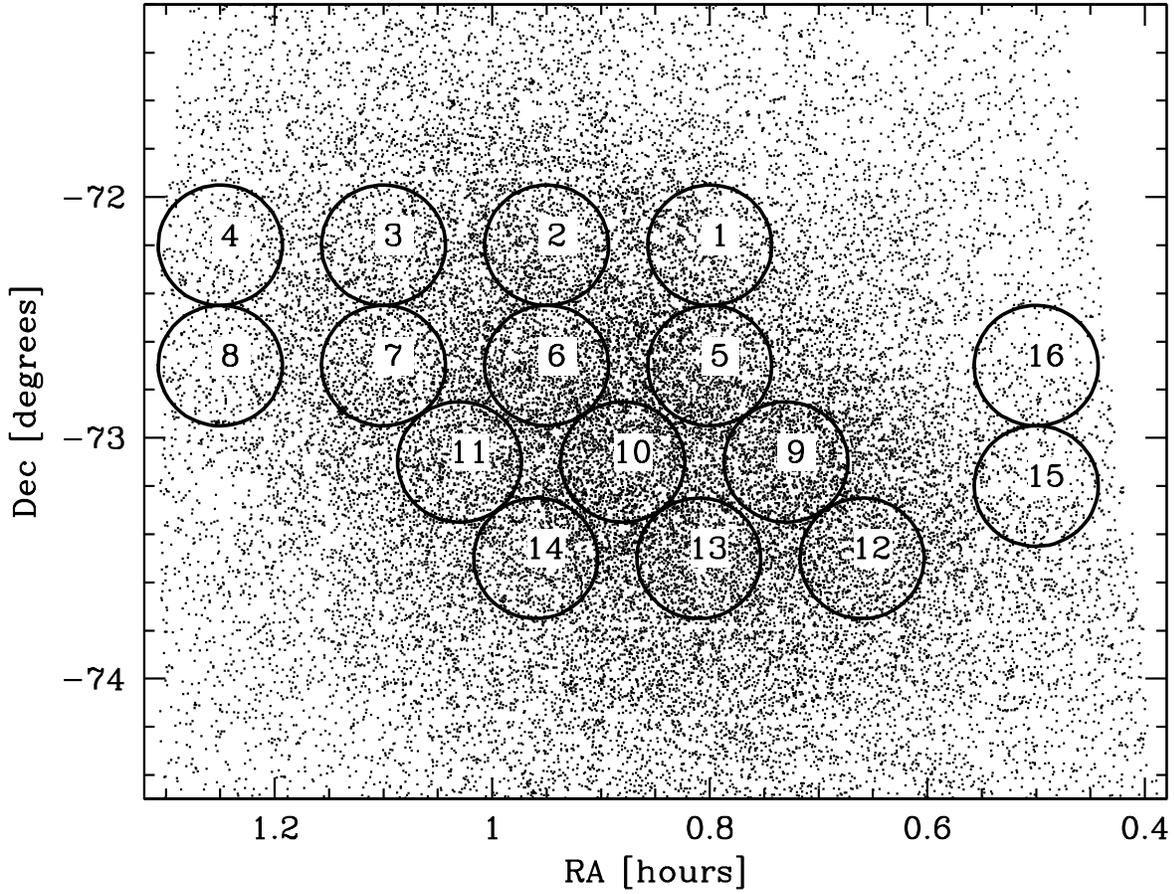}
\caption{ The distribution on the sky of the sixteen IMACS fields
which were prepared for observation.  The points show the 
distribution of 34,000 RGB candidates in the SMC, selected with 
the photometric criteria described in Section~\ref{sec:sample}.
Fields 9 and 10 were not observed. \label{fig:smcfields} }
\end{figure}

We use these photometrically-selected RGB and alignment star samples
as input for the IMACS slitmask design program {\tt intgui}.  Our
sixteen selected fields cover the central 3.9~kpc$\times$1.8~kpc of
the SMC (see Figure~\ref{fig:smcfields}).  In each of these fields, we
interactively select between 7 and 12 alignment stars from the dozens
available to uniformly cover the field of view.  The slit design
package optimally selects objects from the candidate target list to
fill the detector plane with non-overlapping spectral traces.  It
employs an internal model of the optical geometry of IMACS to predict
the spectral trace of each potential target in a given field on the
detector plane.  A target is only selected for observation if its
predicted spectral trace on the detector includes a predefined
wavelength interval of interest.  Because we are targeting the
Ca~{\small II} triplet, we selected a wavelength interval from
8000~\AA\ to 9000~\AA.  After determining the set of objects whose
spectra fall on the detector properly, {\tt intgui} determines whether
there are overlapping traces, and eliminates objects until no overlaps
remain.  One can assign weights to candidate objects to modulate the
likelihood that they are selected for observation, but we chose the
default uniform weighting.  We chose a short slit length (4\arcsec),
to maximize the number of slits per mask.  Our final masks contain an
average of 270 slits (ranging from 122 to 359), which is roughly half
of the available RGB candidates in each field.
Table~\ref{tab:field_data} presents the following data for each
prepared field: (1) the field identification number; (2) the Right
Ascension coordinate of the field center; (3) the Declination
coordinate of the field center; and (4) the number of slits placed on
RGB candidates in the field.  The remaining columns of
Table~\ref{tab:field_data} are described in
Section~\ref{sec:centererrs}.

\subsection{Observations}\label{sec:observations}

We observed 14 of our 16 fields over four half-nights on the Magellan 
Baade telescope, from August 29 to September 1, 2004.  Each night was 
clear, with seeing between 1\arcsec\ and 1.5\arcsec.  Our fields were 
observed at an airmass between 1.4 and 1.5, and the moon was nearly 
full throughout the run.  We used the f/2 camera, which images a 
27.4\arcmin\ diameter field onto the 8k$\times$8k CCD detector array, 
with an image scale of 0.2\arcsec\ per pixel.  We used the 
600~l~mm$^{-1}$ grism blazed at 34$^\circ$, which covers 5600~\AA\  
to 10000~\AA\  in first order, at a resolution of 0.583~\AA\  per 
pixel.  Our slits are 1\arcsec\ wide, so the effective spectral 
resolution is 2.9~\AA\  per resolution element.  We used the Clear 
filter to obtain the widest possible wavelength coverage.

An iterative procedure is required to obtain precision pointing that
ensures that each slit is correctly illuminated by the corresponding
target object.  After pointing the telescope to the field center
coordinates, we obtain a short direct-image exposure of the mask.  We
measure the positional offsets of the 7--12 alignment stars in the
field with respect to the centers of their square alignment holes in
the mask.  We then adjust the telescope pointing and field rotation
angle to correct for these offsets, and take another direct image of
the mask.  The process is repeated until the alignment star offsets
are minimized (this typically requires only two iterations).  There
were two fields (01 and 06) for which we could not simultaneously
center all alignment stars.  We performed the best alignment that we
could following the described procedure, but determined later that
these fields still suffered from non-negligible pointing and
field-rotation errors (see Section~\ref{sec:centererrs}).

The remaining observational details are straightforward. With the 
600~l~mm$^{-1}$ grism, IMACS has a total throughput efficiency of 
about 6\% at 8500~\AA.  For the first field observed (field 04), we 
obtained three exposures of 1200~s.  However, from those images we 
determined that two 1200~s exposures would be sufficient to obtain 
S/N = 25 to 50 for our $I\simeq16$~mag targets.  The remaining fields 
were observed with $2\times1200$~s exposures.  Each observation was 
immediately followed by a pair of flat-field exposures obtained 
without changing the telescope orientation by moving an illuminated 
flat-field screen into the optical path.  We obtained arc-lamp 
exposures of each field in the daytime, which is sufficient for a 
first-order wavelength solution.  We use the night-sky lines embedded 
in the target spectra for our final wavelength calibration (see 
Section~\ref{sec:reduction}).  A series of ten bias exposures was 
also obtained each afternoon.

\subsection{Data Reduction}\label{sec:reduction}

We use IRAF\footnote{IRAF is the Image Reduction and Analysis 
Facility, distributed by the National Optical Astronomy 
Observatories, which are operated by the Association of Universities 
for Research in Astronomy, Inc., under cooperative agreement with the 
National Science Foundation.} and the COSMOS reduction package, the 
latter of which is designed specifically to reduce, calibrate, and 
extract multislit IMACS spectra. COSMOS consists of a collection of 
programs, each of which performs a specific data-reduction task with 
minimal input required from the user.  In the following description 
of our procedure, we designate IRAF tasks as I:{\tt task} and COSMOS 
tasks as C:{\tt task}.

We use I:{\tt zerocombine} to combine our ten daily bias exposures 
into a master bias image for each night.  We then run C:{\tt
Sflats}, which removes the bias from the flat field images, combines
multiple exposures to reject cosmic rays, and normalizes each combined
flat to a mean pixel value of 1.0.  To complete the preliminary image
processing, we run C:{\tt biasflat} on each spectroscopic image to
subtract the bias and divide by the corresponding normalized
flat-field image.

We used a two-step process to simultaneously fit a 2-D spectral trace
to each slit, and determine a preliminary wavelength calibration. We
first ran C:{\tt align-mask}, which uses an internal model of the 
optical geometry of IMACS and the known positions of each slit in a 
given mask to predict the spectral trace of each spectrum on the 
detector plane.  The predicted traces are then refined by matching to 
the arc-lamp spectral images, using the C:{\tt map-spectra} task. 
C:{\tt map-spectra} uses third-order polynomials to fit the tilt,
curvature and width of each trace, and it uses the known emission-line
features in the arc lamp spectra to simultaneously fit a fifth-order
polynomial along the dispersion direction of each trace.

The standard COSMOS procedure is to run C:{\tt subsky} {\it before} 
extracting the spectra, which produces better results since the sky 
pixels are not resampled.  However, we postpone sky subtraction 
because we will use the sky emission lines for our final wavelength 
calibration.  Therefore, we next use C:{\tt extract-2dspec} to apply 
the optimized trace models to the data images, and extract the 
rectified, wavelength-calibrated, two-dimensional spectra from each 
individual field exposure.  We use C:{\tt sumspec} to combine each 
extracted exposure pair and reject cosmic rays (C:{\tt sumspec} 
adopts the average pixel value of the pair, unless the two pixel 
values differ by more than $3\sigma$, in which case it adopts the 
lower value).

The density of emission lines in the observed spectral region 
($7000$~\AA$ < \lambda < 10000$~\AA) is more than sufficient to 
support a robust wavelength calibration.  We define and apply the 
wavelength calibration using the standard IRAF pipeline of 
I:{\tt identify}, I:{\tt reidentify}, I:{\tt fitcoords}, and 
I:{\tt transform}.  I:{\tt identify} is run on a single 
representative slit from each field that is selected to cover a wide 
wavelength range with large S/N.  I:{\tt identify} uses a database of 
spectral features with known wavelengths to determine the wavelength 
solution.  We choose the ``lowskylines'' database, which is included 
with IRAF and is appropriate for night sky lines in a low-resolution 
spectrum. Because the spectra already have a first-order calibration 
from the COSMOS fit to our daily emission lamp spectra, we do not 
need to manually identify any spectral features and instead we allow 
I:{\tt identify} to automatically identify emission features from the 
``lowskylines'' database.  We fit a fourth-order Chebyshev function 
to minimize the deviations of the observed line centers from their 
known positions and obtain a refined wavelength solution.  We reject 
points which deviate from the wavelength solution by $>3\sigma$ as 
outliers and recompute the fit.  We then reidentify features using 
the improved model and iterate the fit until we reach a solution that 
has an RMS $<0.3$~\AA. 

We run I:{\tt reidentify} on all remaining slit images in the field,
using the wavelength calibration found by I:{\tt identify} as a
starting point, and automatically determining an optimal wavelength
solution for each slit.  I:{\tt reidentify} determines an independent
wavelength solution for every fourth row in each two-dimensional slit
image (for a total of five rows per slit), and uses the same fitting
function (fourth order Chebyshev) and set of emission lines as used 
by I:{\tt identify}.  Next, we run I:{\tt fitcoords} to fit a 
two-dimensional surface to the set of five wavelength solutions for 
each slit obtained by I:{\tt reidentify}.  I:{\tt fitcoords} uses a 
second-order Chebyshev function in the dispersion direction and a 
first-order Chebyshev function in the spatial direction to fit the 
surface.  Finally,  we use I:{\tt transform} and the two-dimensional 
wavelength solution to rectify each image.

\begin{figure}[h]
\plotone{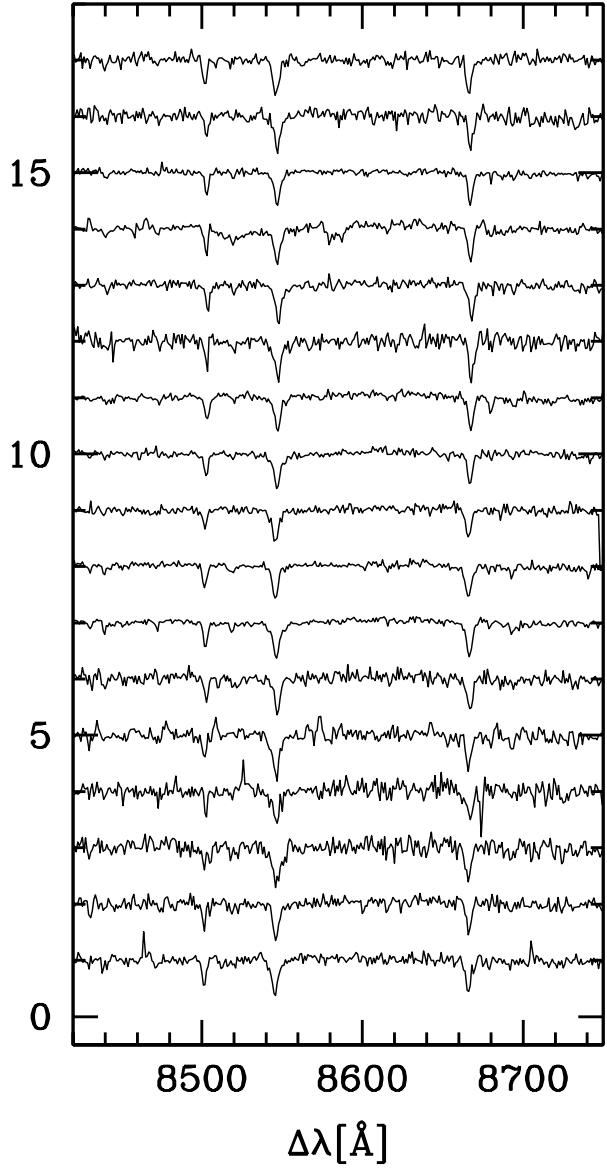}
\caption{ A sampling of extracted spectra from one of our fields
(field 07), focusing on the Ca~{\small II} triplet region. 
\label{fig:spectra}}
\end{figure}

The final reduction step is to extract the one-dimensional spectral
trace of each target star from its rectified slit image and subtract
the sky using the I:{\tt apall} task.  I:{\tt apall} works well in its
non-interactive mode for these data because most of the curvature in
the spectral trace has already been removed by the two-dimensional
COSMOS extraction.  A sample of the extracted one-dimensional spectra
from one of our fields is shown in Figure~\ref{fig:spectra}.

\begin{figure}[h]
\plotone{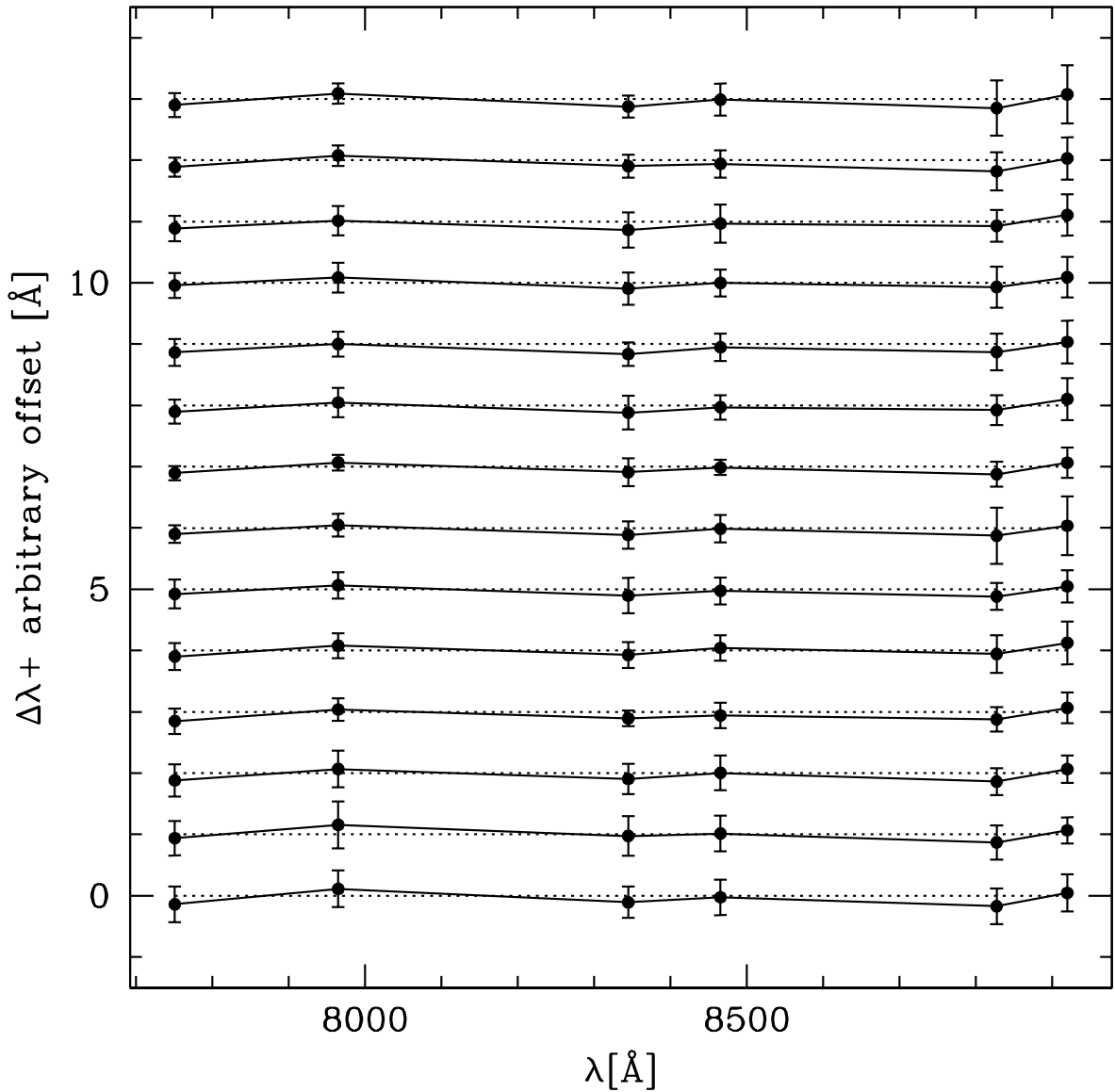}
\caption{Checking for systematic errors in the wavelength calibration.
We plot the deviation of the average measured wavelength for six
bright sky emission lines in 2079 rectified, one-dimensional
extracted spectra.  Each set of connected points shows the results
for one of our 14 observed fields, and the error bars indicate the
standard deviation of the measurements of each line.  The data for
each field are given an arbitrary offset for clarity, and the dotted
lines indicate the zero-deviation line for each field.  While there
are detectable systematic trends that are persistent across all
fields, these deviations are within one standard deviation of the
tabulated wavelength and smaller than the quoted random errors for
individual stars. \label{fig:wavecalib} }
\end{figure}

To check the quality of the wavelength calibration, we select a subset
of prominent sky emission lines and measure their wavelength in the
set of 2079 rectified, 1-dimensional extracted spectra for which sky
subtraction was omitted (see Figure~\ref{fig:wavecalib}).  The
standard deviation of the distribution of measured sky line positions
is between $0.25$~\AA\ and $0.30$~\AA (roughly 0.1 resolution
element).  The mean deviation of the measured line positions is
$0.09$~\AA, corresponding to a velocity uncertainty of 3~km~s$^{-1}$.

\subsection{Measuring Line Profiles}\label{sec:lineprofs}

We measure the central wavelengths and equivalent widths of the
Ca~{\small II} triplet lines at 8498~\AA, 8542~\AA\ and 8663~\AA\ in
each of our red giant spectra to determine the stellar radial velocity
and calcium-triplet metallicity.  Each line is fit independently, with
no {\it a priori} constraint on their relative position or line
strength. We isolate 20~\AA\ segments of the spectrum, centered on the
expected wavelength of each line (assuming a mean radial velocity for
the SMC), and fit a straight-line segment (for the stellar continuum)
plus an inverted Gaussian function (for the line profile) to each
segment.  The fit parameters of the Gaussian yield the central
wavelength and equivalent width of each line.  To estimate the
uncertainty on these quantities, we employ a Monte Carlo technique.
We construct 50 realizations of the spectral segments by perturbing
each pixel's flux value according to its estimated uncertainty.  We
fit for the central wavelength and equivalent width in each of these
realizations, and adopt the standard deviation of their distribution
as the estimated internal uncertainty of our measurements.  

\begin{figure}[h]
\plotone{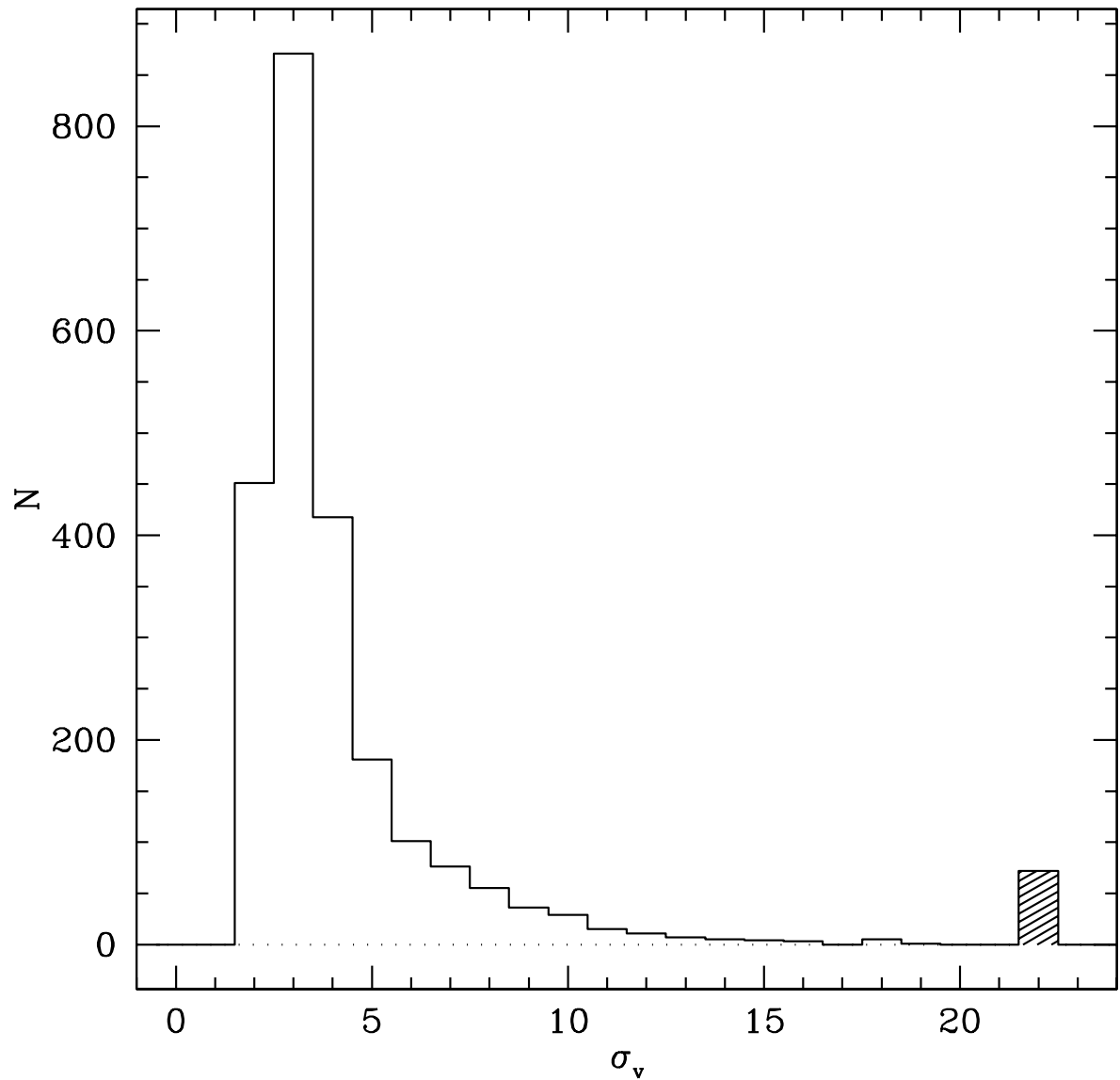}
\caption{The distribution of measured radial velocity uncertainties,
  as derived from the Doppler shift of each spectrum (not including
  uncertainties from the Heliocentric correction or field-rotation
  errors).  The distribution has a narrow peak at 3~km~s$^{-1}$, and
  97\% of the stars have $\sigma_v < 20$~km~s$^{-1}$.  The shaded bin
  at the right indicates the number of stars with $\sigma_v >
  20$~km~s$^{-1}$.  These stars are actually distributed more-or-less
  uniformly between 20 and 120~km~s$^{-1}$, and are compressed into a
  single bin in the plot for clarity. \label{fig:verrs}}
\end{figure}

The radial velocity of each star is determined using the mean Doppler
shift of the three calcium lines relative to their known rest
wavelengths.  The distribution of uncertainties in the measured
doppler shift is very tightly peaked at 0.08~\AA, and 97\% of the
stars have a measurement uncertainty below 0.6~\AA.  However, there is
a long, sparsely-populated tail in the distribution, to uncertainty
values as high as 3~\AA.  Most of the spectra with large uncertainties
are polluted by spectral-overlap artifacts that were not flagged in
our visual inspection of the sample (see below).  We reject objects
whose uncertainty in the Doppler shift of the Calcium lines exceeds
0.6~\AA, corresponding to a maximum allowed radial velocity error of
20~km~s$^{-1}$ (see Figure~\ref{fig:verrs}).

Of the 3650 observed slits, we successfully extracted 3416
one-dimensional spectra.  Of these, 789 were rejected due to large
artifacts in the extracted spectrum, 240 spectra were rejected because
fewer than two Calcium lines could be measured, and another 82 spectra
were rejected because their estimated radial velocity uncertainty
exceeded $20$~km~s$^{-1}$.  This leaves 2305 spectra with measured
radial velocities.

We discovered during our reduction of the data that the 
600~l~mm$^{-1}$ grism was slightly misaligned in its IMACS holding 
cell during our observing run. As a result, there were systematic 
errors in the model of the spectral traces in the detector plane.  
These errors caused some of the spectral traces to overlap on the 
detector, because the slitmask design program {\tt intgui} had 
assumed the wrong model for spectral traces.  The overlapping traces 
produce severe artifacts in the extracted spectra that lead to the 
above rejections.  In total, we lost approximately 30\% of our 
observed spectra to overlap-related artifacts.

\subsection{Systematic Velocity Errors}\label{sec:centererrs}

In slit spectroscopy, there is the potential for systematic velocity
errors introduced by misalignments of the star and slit.  The
magnitude of the error is $\Delta v = c \times \Delta \theta \times p
/ \lambda_0$, where $c$ is the speed of light, $\Delta \theta$ is the
angular offset of the star from the slit center in arcseconds, $p$ is
the spectral resolution in \AA~arcsec$^{-1}$ (2.9 for these data), and
$\lambda_0$ is a characteristic wavelength for the measured features
($\sim8600$~\AA\ for the Ca~{\small II} triplet).  

Applying the above formula in a straightforward manner implies that
the systematic velocity error due to pointing could be as high as
50~km~s$^{-1}$, for a star positioned at the edge of the slit.
However, in practice the real systematic error will never be this
large, because it is mitigated by the finite size of the star's image.
We define an ``effective'' pointing offset, $\Delta \theta_{eff}$,
which is the offset from the slit center of the centroid of the star's
flux distribution in the slit.  Because the seeing was consistently of
order the width of our slits (1\arcsec), the effective worst-case
velocity uncertainty for a star centered at the edge of the slit is
actually around 25~km~s$^{-1}$.  The iterative pointing procedure we
employed (see Section~\ref{sec:observations}) is designed to minimize
such pointing errors, and results in a positioning of the field to
within 0.2\arcsec.  This is also equivalent to the internal
astrometric precision of the MCPS catalog.  Adopting 0.2\arcsec as a
characteristic pointing uncertainty yields a systematic velocity
uncertainty due to pointing errors of 10~km~s$^{-1}$.

\begin{figure}[h]
\plotone{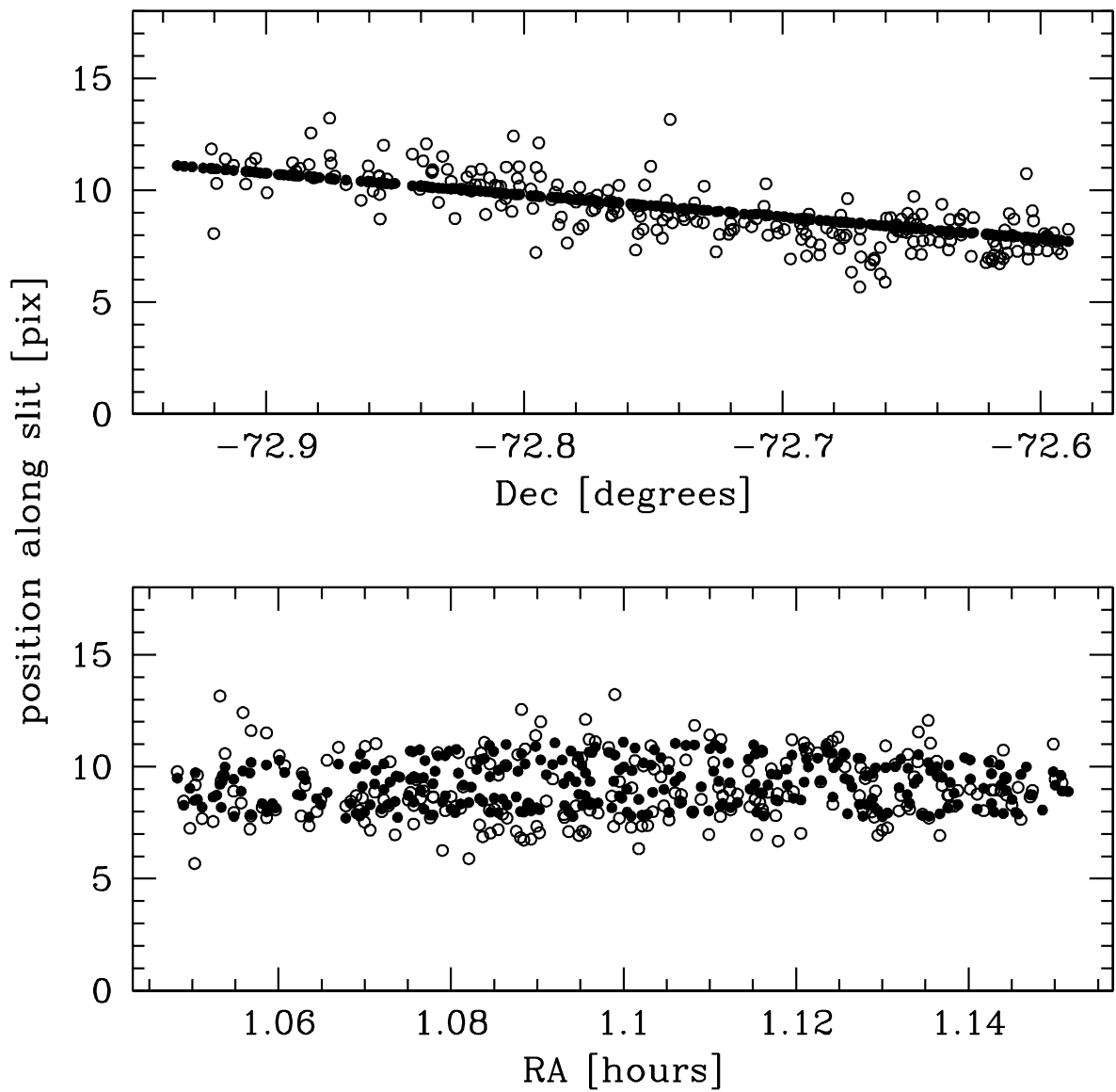}
\caption{ Open points: the measured position of the peak of the 
spectral trace along the spatial dimension, for the slits in 
field~07, vs.\ the slit's Right Ascension (bottom panel) and 
Declination (top panel).  Filled points: the expected spectral-trace 
positions if there was a field-rotation error of -0.45$^\circ$.  We 
construct this plot for each field to constrain the field-rotation 
error, which we use to apply systematic velocity corrections.
\label{fig:frotation} }
\end{figure}

In some of our fields, we detected a systematic radial-velocity
gradient across the field, which we attribute to a small error in the
rotational orientation of the slitmask.  We were able to measure this
error in the field orientation by examining the positions of the
stars' spectral traces along the spatial dimension, because such a
field-rotation error introduces a linear gradient in the
spectral-trace positions in the Declination direction.  Therefore, by
measuring the gradient in the peak positions of the spectral trace as
a function of Declination, we can constrain the field-rotation error.
This is illustrated for one of the fields (Field 07) in
Figure~\ref{fig:frotation}, in which we show the measured peak
position of the spectra in the spatial direction as a function of both
Right Ascension and Declination. As expected, there is no observed
gradient in Right Ascension, and the gradient in Declination is
well-matched by the predicted effect of a rotational error of
$-0.45^\circ$.  We fit the measured peak positions in each field in
this manner to determine the best-fit field-rotation error for each
field.

\begin{figure}[h]
\plotone{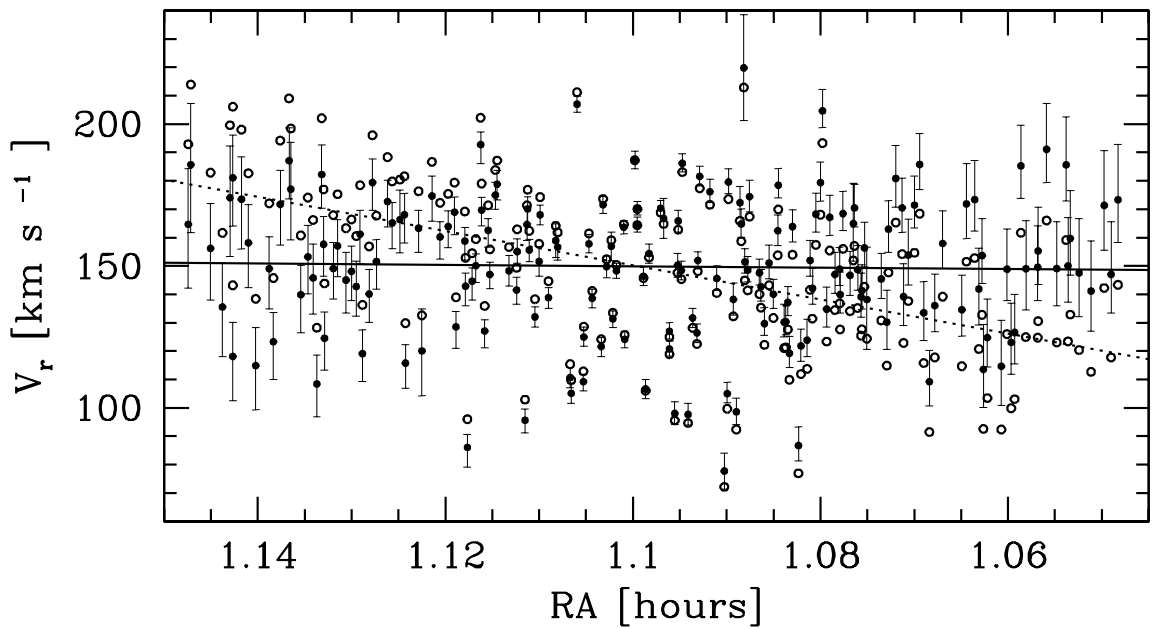}
\caption{ Open points: the measured radial velocities in field 07 as a
function of Right Ascension.  Solid points with errorbars: the radial
velocities, after correcting for the field-rotation error as measured
by the positions of the spectral traces.  The dotted and solid lines
are the best linear fits through the observed and corrected data,
respectively.  These fits reveal that the observed apparent velocity
gradient was in fact an artifact due to the field-rotation error. 
This is found to be the case for each field for which a non-zero 
field-rotation error was measured. \label{fig:vcorr} }
\end{figure}

The field-rotation error also induces a complementary artificial
radial-velocity gradient along the Right Ascension direction in each
field.  In Figure~\ref{fig:vcorr}, we show the measured radial
velocities of stars in Field 07 as a function of Right Ascension (open
circles), and the corrected velocities predicted by our measurement of
the field-rotation correction of $-0.45^\circ$ (filled circles).  In
each field for which we determined a non-zero rotation error, applying
a correction in this manner flattens the observed radial-velocity
gradient, and in most cases reduces the RMS scatter of the velocity
distribution by up to 16\%.  Applying this correction introduces an
additional uncertainty due to uncertainty in the best-fit rotation
angle.  This uncertainty is indicated with the error bars in
Figure~\ref{fig:vcorr}.  The magnitude of the uncertainty depends on
the goodness-of-fit of the field-rotation correction, and also
increases linearly with the offset of a given slit from the field
center, in the Right Ascension direction.  

During the analysis of Fields~01 and 06, we found evidence of large
pointing errors involving both translation and rotation.  The rotation
errors in Field~06 were so large that stars near the top and bottom of
the field lay outside the bounds of their slits.  In both fields, the
mean position of the spectral peaks in the spatial direction was
several pixels away from the center of the slit.  While we cannot
directly measure the mean offset in the dispersion direction, the
error in the spatial direction is cause for concern about the absolute
velocity measurements in these fields. We choose to exclude the 226
stars in these fields from our analysis of the SMC kinematics.
However, we note that the data from Fields~01 and 06 will be included
in Paper~II, since the metallicity analysis depends only on the
equivalent widths of the Calcium lines, and not on their wavelengths.

The coordinates and velocities for the final sample of 2079 stars for
which we measured velocities are presented in Table~\ref{tab:smcrgb}.
The observed (topocentric) radial velocities are corrected to the
heliocentric reference frame, using the IRAF task {\tt rvcorrect}.
The columns of Table~\ref{tab:smcrgb} are: (1) an identification
number for the object; (2) the Right Ascension coordinate; (3) the
Declination coordinate; (4) the $V$ magniutde; (5) the $I$ magnitude;
(6) the observed topocentric radial velocity; (7) the radial velocity,
corrected to the Heliocentric frame; (8) the 1-$\sigma$ uncertainty in
the positive $v_r$ direction; and (9) the 1-$\sigma$ uncertainty in
the negative $v_r$ direction.  The estimated radial-velocity
uncertainties in columns 8 and 9 are usually dominated by the error
due to the pointing precision of 0.2\arcsec.  They also include
uncertainties associated with the field-rotation correction (see
Figure~\ref{fig:vcorr}), internal uncertainties from our measurement
of the Ca~{\small II} triplet lines, and the uncertainty of the
wavelength calibration (see Figure~\ref{fig:wavecalib}).  The
remaining discussion employs the corrected heliocentric radial
velocities.

In Table~\ref{tab:field_data}, we present additional data for the
twelve fields which were observed.  The additional columns are: (5)
the number of RGB stars with a measured radial velocity; (6) the mean
radial velocity from a best-fit Gaussian; (7) the velocity dispersion
from a best-fit Gaussian; (8) result of a K-S test investigating
variations in the velocity distributions (see
Section~\ref{sec:kinematics}); (9) the best-fit field-rotation
correction angle; and (10) the mean radial-velocity uncertainty due to
the field-rotation correction.

\subsection{Comparison to Previous Measurements}\label{sec:compare}

When designing our target sample, we specifically set out to include
red giant stars in the SMC whose radial velocities had been determined
by previous authors.  Unfortunately, very few such studies exist, and
we found only one, \citet{dh98}, that reported observations of
individual SMC giants.  Of the seven SMC clusters targeted in that
study, only Lindsay~11 falls within the bounds of the MCPS.  We
observed three of the six red giants in Lindsay~11 that were observed
by \citeauthor{dh98} (MJD~182, MJD~212, and MJD~240, see
Table~\ref{tab:smcrgb}).  While \citeauthor{dh98} report the mean
cluster velocity for Lindsay~11 (132$\pm$5~km~s$^{-1}$), they do not
report the individual velocities of the six giants they observed.  The
mean velocity of our three Lindsay~11 stars is 137$\pm$5~km~s$^{-1}$,
in good agreement with the \cite{dh98} result.

\subsection{Removing Foreground Contaminants}\label{sec:foreground}

We expect our photometrically-selected sample to include a small
number of foreground Milky Way main sequence dwarfs.  Many of these
contaminants can be identified by their radial velocity.  However, 
some overlap in radial velocity is inevitable, so a second criterion 
is needed to cleanly remove the foreground population.  We employ the 
Na doublet absorption feature at (8183~\AA\ and 8195~\AA), which is 
sensitive to surface gravity (\ie, it is strong in dwarfs, and weak 
or absent in giants).

\begin{figure}[h]
\plotone{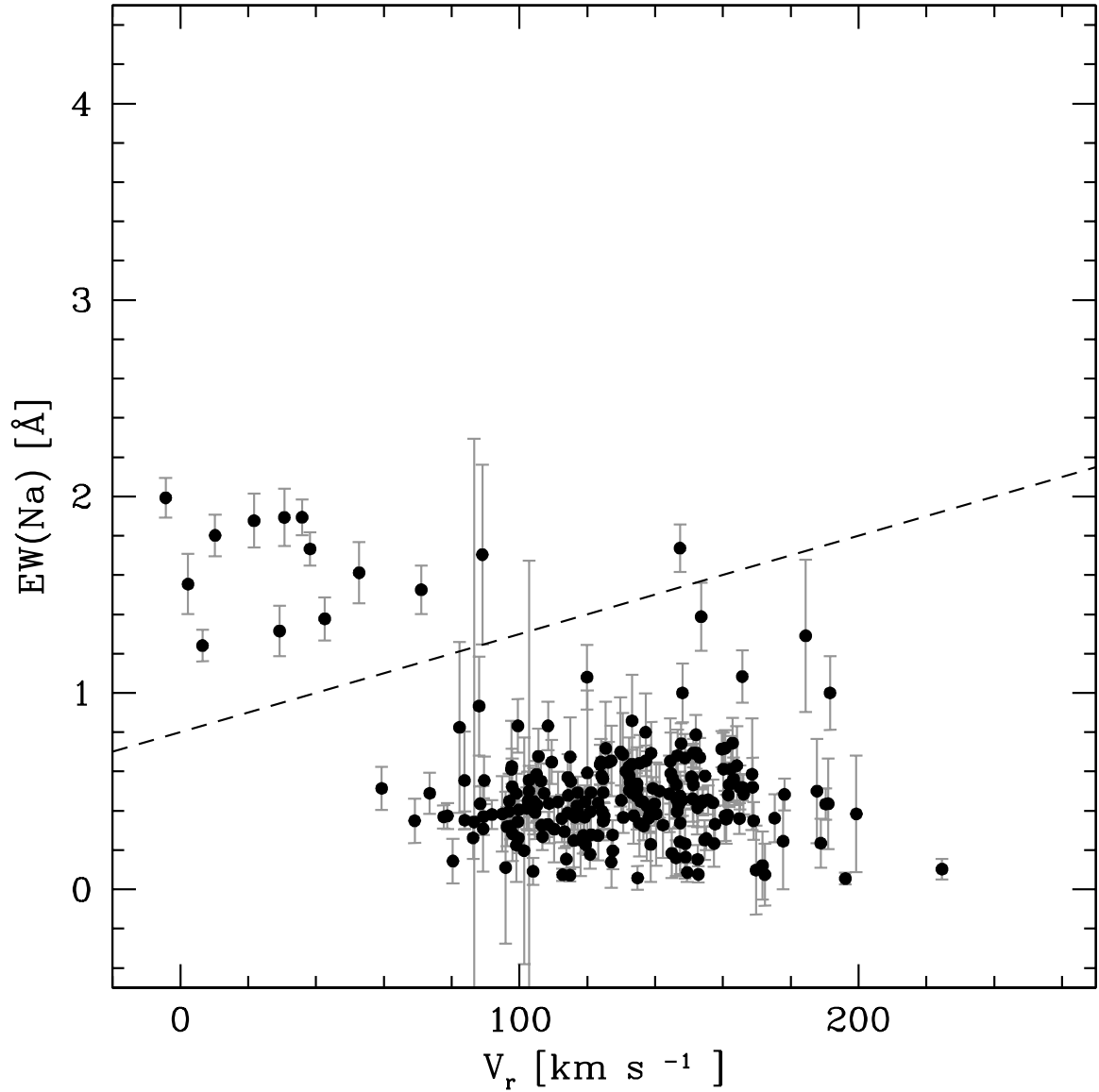}
\caption{ The total equivalent width of the Na doublet vs. the radial
velocity of the star, for the subset of stars in which the Na doublet
was detected.  Foreground dwarfs in our sample will have low 
velocities and large EW(Na).  Objects above the dashed line are 
identified as foreground stars in our sample.\label{fig:nadoublet} }
\end{figure}

In Figure~\ref{fig:nadoublet}, we plot radial velocity against the
equivalent width of the Na doublet for the subset of stars in our
sample in which the Na doublet was measured.  The foreground
population is  the set of points with low velocity and large EW(Na). 
However, there is a small population of objects that have large
EW(Na), but also have large velocities.  We visually inspected the 
objects for which EW(Na)$\simgreat$1 and $v_r\simgreat50$~km~s$^{-1}$, 
finding that some of the large measured EW(Na) values were spurious, 
or affected by obvious artifacts.  The dotted line in 
Figure~\ref{fig:nadoublet} indicates our adopted separation between 
dwarfs and giants.  In addition to the 14 foreground stars shown in 
Figure~\ref{fig:nadoublet}, 14 more objects are flagged as foreground 
objects, solely on the basis of their SMC-discrepant radial 
velocities ($v_r < 50$~km~s$^{-1}$ or $v_r > 250$~km~s$^{-1}$).  
EW(Na) could not be measured properly in these objects, usually due 
to extraction artifacts or the gap between CCDs in the detector 
array.  The 33 stars determined to be unassociated with the SMC are 
flagged in Table~\ref{tab:smcrgb}, and are omitted from the remaining 
discussion.

\section{SMC Kinematics and Dynamical Structure}\label{sec:kinematics}

\begin{figure}[h]
\plotone{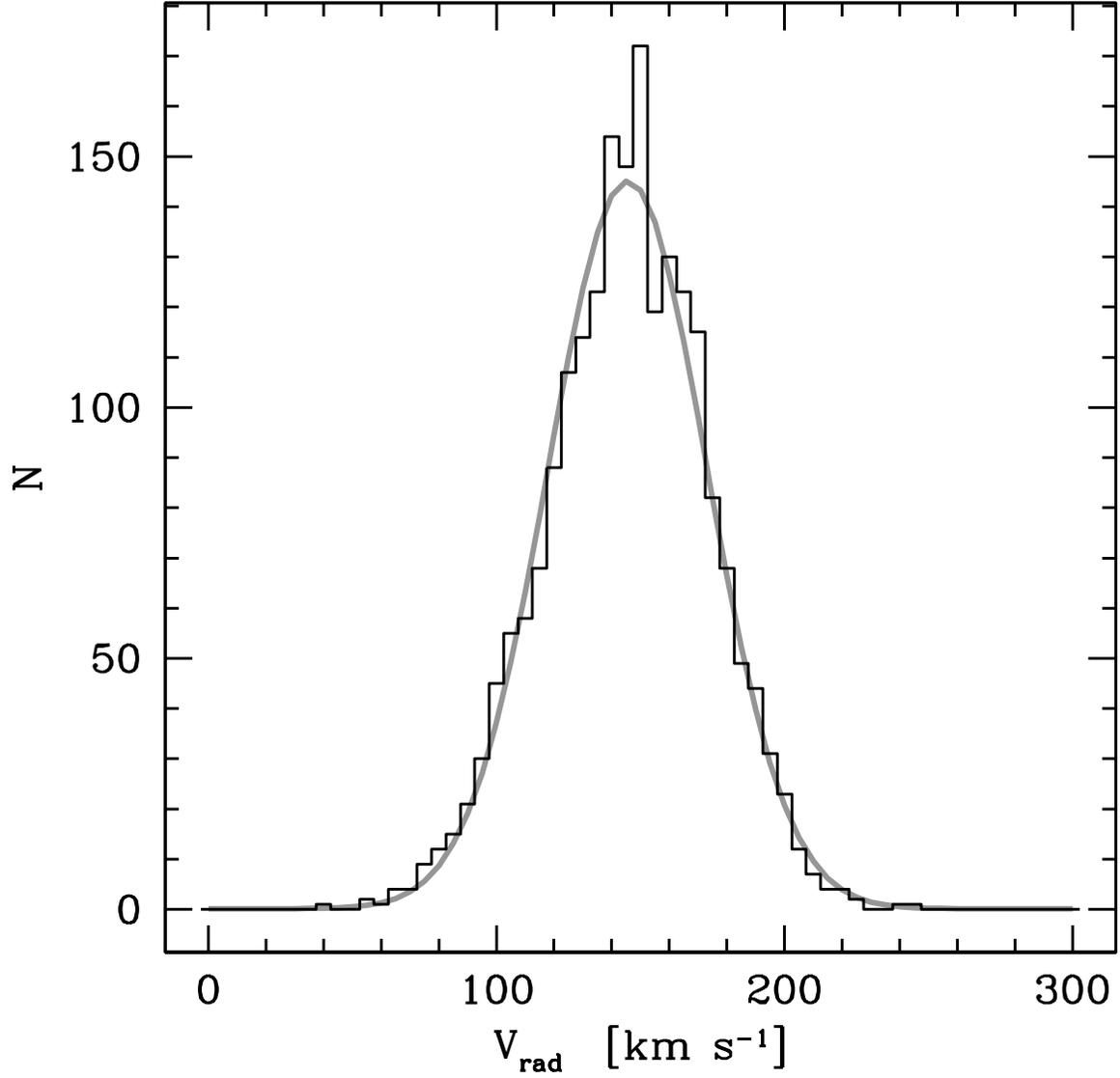}
\caption{ Histogram: the distribution of radial velocities of 2046 red
giants in the SMC.  It is well-fit by a Gaussian (grey curve) centered
at 146~km~s$^{-1}$, with a dispersion of 28~km~s$^{-1}$.
\label{fig:vrad_hist} }
\end{figure}

In Figure~\ref{fig:vrad_hist}, we show the global distribution of
radial velocities of the 2046 SMC stars in our sample.  It is
evidently well-fit by a Gaussian, which has a mean velocity of
$145.6\pm0.6$~km~s$^{-1}$ and a velocity dispersion of
$27.6\pm0.5$~km~s$^{-1}$.  This compares well with previous
radial-velocity studies of the SMC, as shown in
Table~\ref{tab:rvsamples}.  It is interesting to note that all
kinematic studies of the SMC show rough agreement in their mean
heliocentric radial velocity ($\sim150$~km~s$^{-1}$) and velocity
dispersion ($\sim25$~km~s$^{-1}$), regardless of the age of the
objects sampled, from H\small{I} gas shells to red giant stars.

A more detailed inspection of our velocity distribution reveals that
the mean velocity of stars in each field varies significantly, from
129 to 156~km~s$^{-1}$ (see Table~\ref{tab:field_data}).  K-S tests
indicate that while most of the fields are consistent with having been
drawn from the same parent distribution, there are some statistically
significant (probability of random $< 10^{-3}$) exceptions: fields 02,
05, 08 and 13.  To investigate whether these field-to-field variations
represent a global kinematic trend, such as systemic rotation or
viewing angle dependent effects, we fit a plane to the velocity
distribution on the sky. Fitting a plane is a simple first order
model, and can represents either solid body rotation or velocity
gradients caused by the projection of differing amounts of the
Galactocentric tangential SMC velocity onto the line-of-sight toward
one side or the other of the SMC. A fit to all of the stars in the
sample results in an excellent fit ($\chi^2 = 1.15$) with an apparent
rotation amplitude of 8.3~km~s$^{-1}$~deg$^{-1}$ and a rms scatter of
$27.5\pm0.5$~km~s$^{-1}$. The apparent kinematic major axis, that of
the largest velocity gradient, is at 23.4$^\circ$ East of North. The
skewness of the velocity distribution is -0.11 and the excess kurtosis
(above that for a Gaussian) is 0.09. The quality of the fit, as
quantified by the low $\chi^2$ value, and the low amplitude of higher
order moments demonstrates that with these data there is no
justification to proceed to more complex fitting models.

The interpretation of the observed velocity gradient is complicated by
the large angular extent of the galaxy on the sky. Because of the
SMC's large size, the tangential velocity of the SMC is projected with
different sign on the two sides of the galaxy, thereby creating an
apparent rotation signature (while the radial velocity is projected
with the same sign). Correcting for this effect requires a
high-precision measurement of the SMC's proper motion. The most recent
measurement of the SMC's proper motion still has uncertainties that
are $>$50\% \citep{kroupa}. A more precise measurement, with an
uncertainty $\sim$5\%, has recently been presented for the LMC
\citep{vdm}.  The latter translates to a tangential velocity of
493~km~s$^{-1}$ for the LMC (the earlier SMC measurement also
suggested a tangential velocity of $\sim$500~km~s$^{-1}$).  If we
adopt a tangential velocity of 500~km~s$^{-1}$ for the SMC, and allow
its direction to be a free parameter, then we can place limits on the
effect this motion has on the magnitude of our measurement of a
rotation velocity by assuming the full effect of the projected
tangential motion is either parallel or antiparallel to that observed.
The adopted tangential velocity will result in an apparent velocity
gradient of 8.7~km~s$^{-1}$~deg$^{-1}$ over the face of the SMC. This
value suggests that the entire velocity gradient observed could be
accounted for by this effect, so it is possible that there is no
intrinsic rotation in the SMC at all. If the result of the projected
tangential motion is oriented in the opposite direction, then the true
velocity gradient due to rotation could be as much as
17~km~s$^{-1}$~deg$^{-1}$. In the direction for which the gradient is
measured, the regions of the SMC that was observed extends about
1$^\circ$ and hence the maximum rotation velocity over the observed
region would be about 17~km~s$^{-1}$.  Even adopting this largest
correction, the rotation velocity is still significantly less than the
velocity dispersion: $v/\sigma \leq 0.6$, so we can conclude that the
SMC is primarily supported by its velocity dispersion.

\subsection{Morphology and Dynamics}

The connection between the shapes of spheroids and their internal
kinematics has been investigated for both giant and dwarf ellipticals
\citep[see][]{davies,dze}. Low luminosity systems generally agree 
with the expectations drawn from models of oblate rotators: their 
degree of flattening is consistent with the expectation of rotational 
support and an isotropic velocity dispersion tensor. We test whether 
the SMC exhibits the same behavior.

\begin{figure}[h]
\plotone{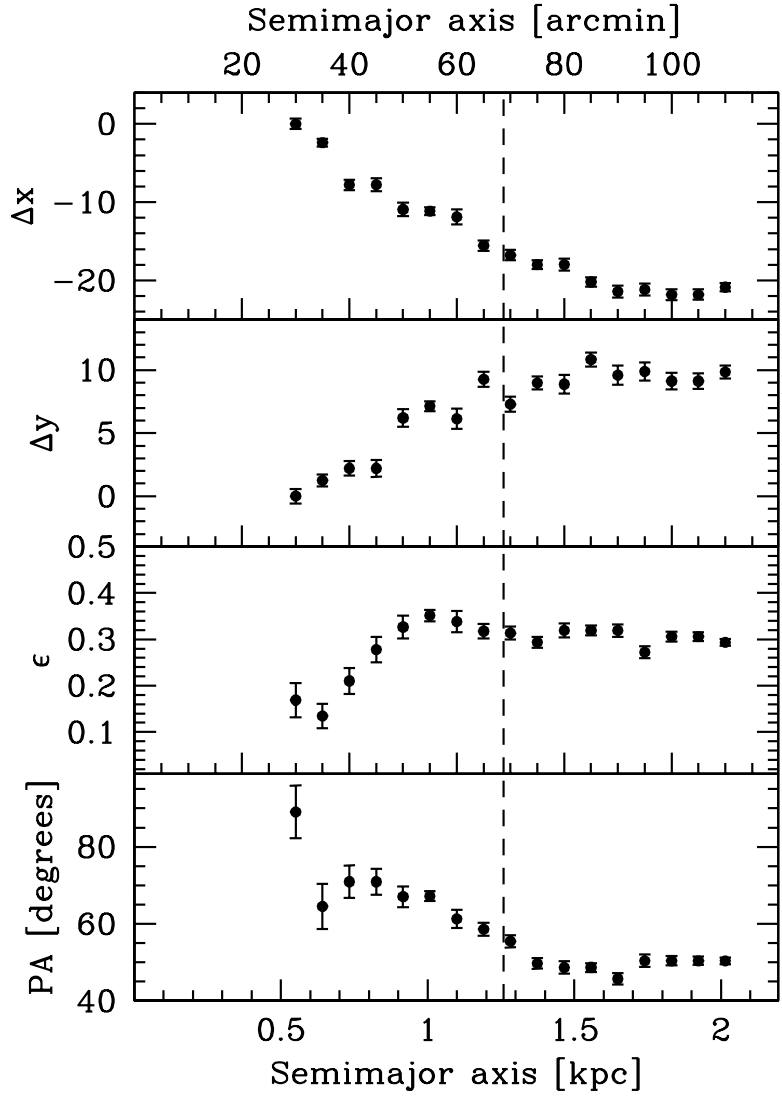}
\caption{Isophote centroid position offset, ellipticity ($\epsilon$), 
and position angle radial profiles. Centroid offsets are in units of 
100\arcsec.
\label{fig:isophotes} }
\end{figure}

We use the stellar catalog presented in \cite{zar02}, and their same
photometric cuts ($16 \le V \le 19.5$, $B-V > 0.7$) to isolate the
older stellar population in the SMC. A stellar density image of the
SMC corresponding to this population is created by doing a gnomic
projection of the coordinates, and accumulating the flux of these
stars in 100\arcsec\ ``pixels''. This image is then analyzed using the
IRAF ellipse fitting routines and the results are plotted in Figure
\ref{fig:isophotes}.  The top two panels show the positional offset of
the center of each elliptical isophote, relative to the innermost
ellipse (with a radius of 30\arcmin), in units of 100\arcsec.  The
third panel shows the ellipticity ($e = 1-b/a$) of each ellipse, and
the fourth shows the position angle of each ellipse's major axis
(measured to the East from North).

The profiles shown in Figure \ref{fig:isophotes} have several features
that bear discussion. First, we have excluded the inner region ($r <
30^\prime$) because of crowding-induced incompleteness in the
photometric catalog. Second, beyond that radius the position angle of
isophotes varies by about 20$^\circ$, possibly indicating a triaxial
stellar distribution that is not viewed along any of the principal
axes.  However, it is possible that the incompleteness effects extend
beyond $30\arcmin$, and these effects could potentially cause the
observed isophotal twisting.  Third, at radii beyond $70^\prime$, the
position of the centroid, the ellipticity, and the isophotal position
angle remain constant.  The constant position angle suggests that at
these radii we are viewing the stellar distribution nearly along a
principal axis, and that therefore the ellipticity reflects the true
axis ratio.  For radii $>70^\prime$, the mean values of the
ellipticity and position angle (measured from North to East) are 0.30
and 49$^\circ$, respectively.  These values are in rough agreement with 
measurements by \cite{dev55}, who found $e=0.58$ and 
$PA=45^\circ\pm2^\circ$.

In Figure \ref{fig:vsigma} we show how the ellipticity and $v/\sigma$
compare to other spheroids. We have taken the data from \cite{davies}
for comparison and plotted the relationship for an oblate rotator with
an isotropic velocity dispersion tensor. The shaded area represents
where the SMC lies depending on the nature of the correction due to
the unknown transverse velocity. The shape and internal kinematics of
the SMC are in the range observed for dwarf ellipticals. Its position
in this diagram demonstrates that its shape is consistent with
rotational flattening.

\begin{figure}[h]
\plotone{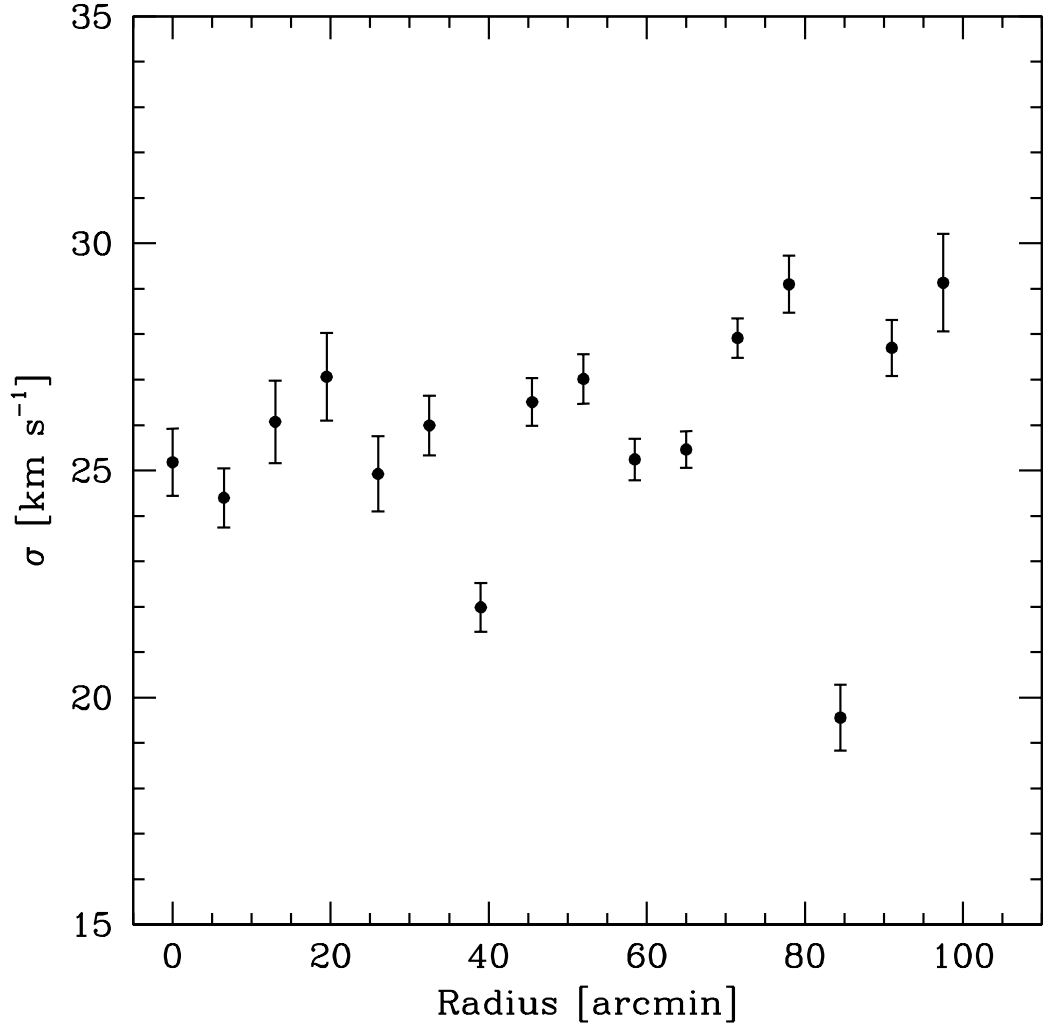}
\caption{Velocity dispersion vs semimajor axis. The velocity
dispersions are calculated using all stars within the elliptical
annulus. The velocities used have been corrected for the linear
velocity gradient.
\label{fig:sigmar} }
\end{figure}

Using the morphological parameters, we define elliptical annuli in
which we evaluate the velocity dispersion (Figure \ref{fig:sigmar}).
The velocity dispersion is roughly flat as a function of radius, with
perhaps a slight rise with increasing radius. Once again we find no
signature of disturbance or deviations from the expectations for a
relaxed spheroidal system.

\subsection{Mass}

Using a highly simplified application of the tensor virial theorem
(assuming a steady-state SMC, an isotropic velocity dispersion, no
projection correction to the rotational velocity, and the potential
energy of a uniform sphere),

$$v^2 + \sigma^2 = {{3GM}\over{5R}},$$

\noindent
we estimate the mass of the SMC out to our last measured point, 
1.6~kpc. For a line-of-sight velocity dispersion of 27.5~km~s$^{-1}$,
which for the assumption of isotropy implies a true velocity
dispersion of 48~km~s$^{-1}$, and a rotation velocity that is
somewhere between 0 and 17~km~s$^{-1}$, we derive that the enclosed
mass is between $1.4\times10^9 M_\odot$ and $1.9\times10^9 M_\odot$. 
In order to compare to the mass measurement obtained by \cite{ssj04}
from the H{\small I} rotation curve, we can extend the analysis to 
3~kpc by assuming that the velocity dispersion remains constant, and 
that the velocity gradient can be extrapolated.  In this case, we 
derive an enclosed mass between $2.7\times10^9 M_\odot$ and 
$5.1\times10^9 M_\odot$. This range of values is slightly larger than 
the mass determined from the H{\small I} rotation curve over the same 
radius \citep[$2.4\times10^9 M_\odot$;][]{ssj04}, but consistent 
given the uncertainties and gross simplifications in our analysis. A 
precise value for the proper motion for the SMC will warrant a more 
complete dynamical treatment and is also necessary to examine higher 
order moments of the velocity field.

\section{Summary}\label{sec:summary}

We present the first spectroscopic survey of field red giant stars
over the main body of the Small Magellanic Cloud.  Our spectroscopic
sample includes over 3000 red giants in 14 28$^\prime$ fields spanning
the central $2^\circ\times1^\circ$ of the SMC.  After rejecting two
fields due to unresolved pointing errors and losing additional spectra
to overlap issues, our final kinematic sample contains 2079 stars
throughout the central 3.9~kpc$\times$1.8~kpc, of which 2046 are
determined to be SMC members.  The global velocity distribution is
well-fit by a Gaussian centered at 146~km~s$^{-1}$ with a width or
$\sigma$ of 28~km~s$^{-1}$, but we also find evidence for a velocity
gradient across the SMC.  Removing the gradient using a planar fit to
the velocities results in an rms scatter ($27.5\pm0.5$~km~s$^{-1}$)
that is effectively unchanged from the global velocity dispersion
measured prior to any correction because the amplitude of the fitted
rotation gradient (7.9~km~s$^{-1}$~deg$^{-1}$) is small. The origin of
the velocity gradient could be either internal rotation or an illusion
caused by the differential projection of the SMC's tangential velocity
along non-parallel lines of sight. Without a high precision
measurement of the SMC's proper motion, a sufficiently precise
correction required to recover the SMC's intrinsic rotation speed is
impossible.  However, even adopting the maximum correction we find
that $v < \sigma$ and hence that the SMC is primarily supported by its
velocity dispersion.

\begin{figure}[h]
\plotone{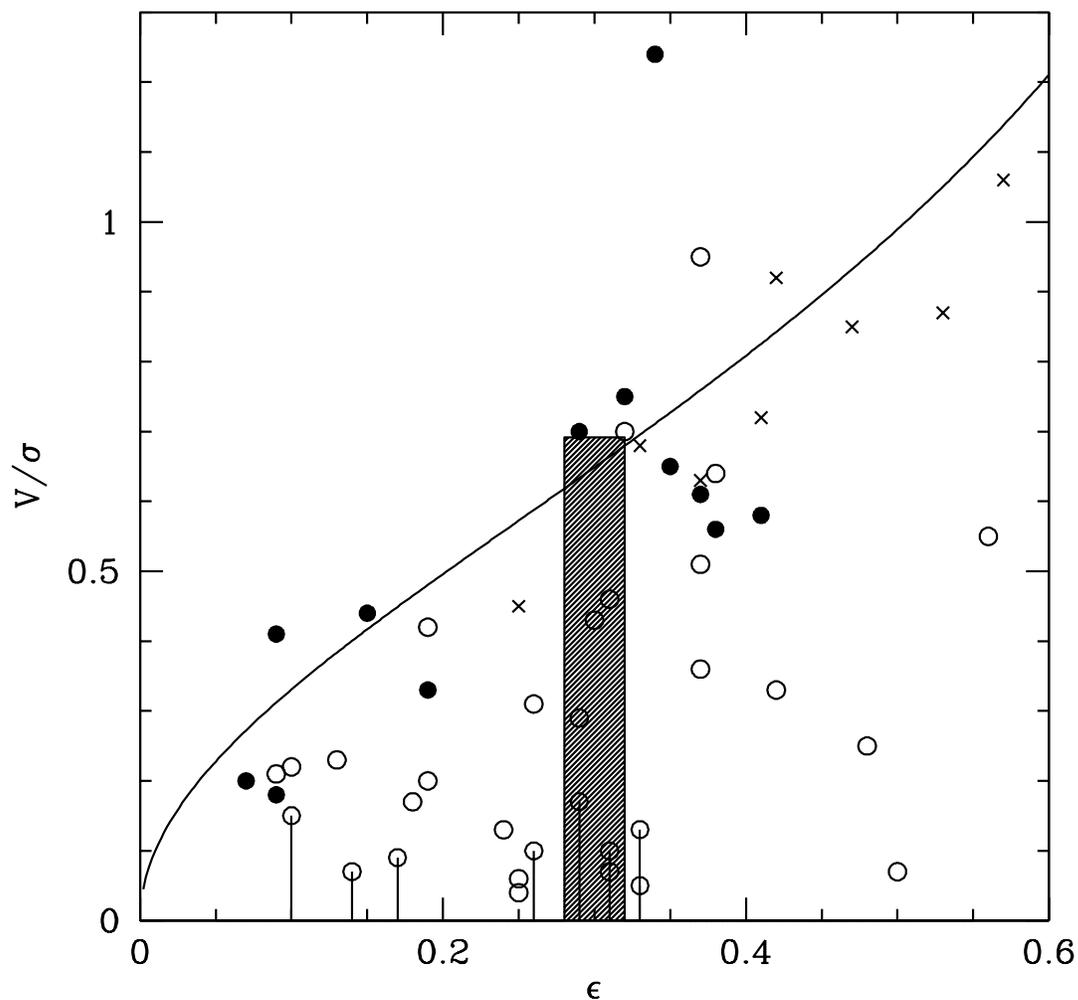}
\caption{The effect of rotation support. We show the distribution of
ellipticity ($1-b/a$) vs. the ratio of rotational velocity to 
velocity dispersion. We plot the original data from \cite{davies}: 
low-luminosity ($M_B > -20.5$~mag) elliptical galaxies are shown with 
filled circles, and bright ($M_B < -20.5$~mag) elliptical galaxies 
are shown with open circles.  Bulges of disk galaxies are shown with 
crosses.  The expected relation for oblate isotropic rotators is 
shown with a solid curve.  The range of values we derive as 
appropriate for the SMC is indicated by the shaded box. 
\label{fig:vsigma} }
\end{figure}

We find that the shape of the SMC, as measured from the analysis of
the spatial distribution of red giant stars, is consistent with the
degree of rotational flattening expected for the observed $v/\sigma$
(Figure \ref{fig:vsigma}). As such, the properties of the SMC are
consistent with similar low luminosity spheroidal systems. We conclude
that the SMC is a low luminosity spheroid whose visual appearance is
dominated by star formation resulting from a recent accretion of gas
\citep[see also][]{zar00}. The underlying morphology and dynamics are
consistent with those of low luminosity ellipticals.

A simple virial analysis using the measured kinematics implies an
enclosed mass within 1.6 kpc of between 1.4 and 1.9$\times 10^9
M_\odot$, and a less well constrained mass of between 2.7 and
$5.1\times 10^9 M_\odot$ within 3 kpc. These values are larger than
those inferred from the H{\small I} kinematics \citep[$2.4\times10^9
M_\odot$,][] {ssj04}, but the simplifying assumptions used could
easily account for differences at the factor of two level.

\vskip 1in
\noindent Acknowledgments:
The authors thank Gus Oemler for his generous assistance in 
getting the COSMOS software working with the misaligned 
600~l~mm$^{-1}$ grating; John Moustakas for providing an IDL 
program for robust spectral line fitting; and J.~D. Smith for 
general IDL assistance.
JH is supported by NASA through Hubble Fellowship grant HF-01160.01-A 
awarded by the Space Telescope Science Institute, which is operated 
by the Association of Universities for Research in Astronomy, Inc., 
under NASA contract NAS 5-26555.  DZ acknowledges financial support 
from National Science Foundation CAREER grant AST-9733111 and a 
fellowship from the David and Lucile Packard Foundation.

\clearpage

\begin{deluxetable}{rrrrrrrrrr}
\tabletypesize{\scriptsize}
\tablecolumns{8}
\tablewidth{0pt}

\tablecaption{Targeted IMACS Fields \label{tab:field_data}}
\tablehead{
    \colhead{ID} & \colhead{R.A.} & \colhead{Dec.} & \colhead{$N_{slits}$} & \colhead{$N_{RGB}$} & 
        \colhead{$<v_{rad}>$} & \colhead{$\sigma$} & \colhead{KS prob.} & 
        \colhead{$\theta_{err}$} & \colhead{$<\sigma_{v,\theta}>$} \\
    \colhead{} & \colhead{[$h:m:s$]} & \colhead{[$d:m$]} &
        \colhead{} & \colhead{} & \colhead{[km~s$^{-1}$]} &
        \colhead{[km~s$^{-1}$]} & \colhead{} & \colhead{[$^\circ$]} & \colhead{[km~s$^{-1}$]} 
}

\startdata  
01\tnm{a} & 0:03:12 & $-$72:12 & 255 &\nodata&  \nodata  &  \nodata & \nodata         & \nodata & \nodata \\
02        & 0:03:48 & $-$72:12 & 293 &   190 & 129$\pm$2 & 28$\pm$2 &$2\times10^{-8}$ &    0.00 & 1.0     \\
03        & 0:04:24 & $-$72:12 & 256 &   160 & 147$\pm$3 & 26$\pm$4 & 0.88            &    0.00 & 2.5     \\
04        & 0:05:00 & $-$72:12 & 146 &   115 & 151$\pm$3 & 20$\pm$2 & 0.06            &    0.00 & 1.4     \\
05        & 0:03:12 & $-$72:42 & 341 &   223 & 136$\pm$2 & 27$\pm$2 &$7\times10^{-4}$ & $-$0.10 & 7.6     \\
06\tnm{a} & 0:03:48 & $-$72:42 & 351 &\nodata&  \nodata  &  \nodata & \nodata         & \nodata & \nodata \\
07        & 0:04:24 & $-$72:42 & 282 &   189 & 151$\pm$2 & 21$\pm$2 & 0.03            & $-$0.45 & 7.0     \\
08        & 0:05:00 & $-$72:42 & 159 &   123 & 156$\pm$2 & 23$\pm$2 &$5\times10^{-4}$ & $-$0.15 & 4.6     \\
09\tnm{b} & 0:02:55 & $-$73:05 & 359 &\nodata&  \nodata  &  \nodata & \nodata         & \nodata & \nodata \\
10\tnm{b} & 0:03:31 & $-$73:05 & 346 &\nodata&  \nodata  &  \nodata & \nodata         & \nodata & \nodata \\
11        & 0:04:07 & $-$73:06 & 304 &   193 & 148$\pm$2 & 25$\pm$2 & 0.60            & $-$0.12 & 8.6     \\
12        & 0:02:38 & $-$73:30 & 326 &   218 & 146$\pm$2 & 28$\pm$2 & 0.90            & $-$0.30 & 6.1     \\
13        & 0:03:14 & $-$73:30 & 341 &   200 & 154$\pm$2 & 25$\pm$2 &$1\times10^{-3}$ & $-$0.25 & 3.8     \\
14        & 0:03:50 & $-$73:30 & 300 &   205 & 149$\pm$2 & 27$\pm$2 & 0.72            & $-$0.25 & 5.5     \\
15        & 0:02:00 & $-$73:12 & 174 &   131 & 141$\pm$3 & 21$\pm$3 & 0.10            & $-$0.35 & 4.4     \\
16        & 0:02:00 & $-$72:42 & 122 &    99 & 148$\pm$3 & 20$\pm$4 & 0.83            & $-$0.50 & 2.2     \\
\enddata
\tablenotetext{a}{Field discarded due to pointing errors}
\tablenotetext{b}{Field not observed}
\end{deluxetable}

\begin{deluxetable}{rrrrrrrrr}
\tabletypesize{\scriptsize}
\tablecolumns{9}
\tablewidth{0pt}

\tablecaption{Red Giant Stars in the SMC \label{tab:smcrgb}}
\tablehead{
    \colhead{ID} & \colhead{R.A.} & \colhead{Dec.} & 
        \colhead{$V$} & \colhead{$I$} & \colhead{$v_{r,obs}$} &
        \colhead{$v_{r,corr}$} & \colhead{+$\sigma$} & \colhead{-$\sigma$} \\
    \colhead{} & \colhead{($h:m:s$)} & \colhead{($d:m:s$)} &
        \colhead{(mag)} & \colhead{(mag)} & \colhead{(km~s$^{-1}$)} &
        \colhead{(km~s$^{-1}$)} & \colhead{(km~s$^{-1}$)} & \colhead{(km~s$^{-1}$)} 
}

\startdata
s02\_0011 & 00:55:39.6 & $-$72:15:59 & 17.24 & 16.11 & 134.1 & 128.7 &  3.2 &  3
.0 \\
s02\_0015 & 00:54:06.1 & $-$72:07:31 & 17.07 & 15.85 & 113.7 & 108.3 &  3.5 &  3
.1 \\
s02\_0018 & 00:54:46.4 & $-$72:06:08 & 17.29 & 16.14 & 157.0 & 151.7 &  4.3 &  4
.1 \\
s02\_0023 & 00:55:22.6 & $-$72:07:05 & 17.18 & 16.03 & 120.0 & 114.7 &  6.1 &  5
.9 \\
s02\_0031 & 00:57:04.5 & $-$72:15:26 & 17.07 & 15.89 & 157.5 & 152.2 &  2.5 &  2
.3 \\
s02\_0035 & 00:57:38.7 & $-$72:20:18 & 16.89 & 15.62 & 112.3 & 107.0 &  3.1 &  2
.9 \\
s02\_0039 & 00:58:58.3 & $-$72:20:39 & 17.34 & 16.20 & 129.7 & 124.4 &  2.9 &  2
.6 \\
s02\_0040 & 00:59:12.5 & $-$72:17:59 & 17.01 & 15.82 &  97.7 &  92.4 &  2.6 &  2
.3 \\
s02\_0046 & 00:58:20.7 & $-$72:11:24 & 17.12 & 15.94 & 105.3 & 100.0 &  2.5 &  2
.4 \\
s02\_0048 & 00:59:02.6 & $-$72:07:16 & 17.23 & 16.07 & 164.3 & 159.1 &  3.7 &  3
.5 \\
\enddata

\tablenotetext{a}{Member of Lindsay~11 observed by \cite{dh98}}
\tablenotetext{b}{Foreground star}
\end{deluxetable}

\begin{deluxetable}{rccl}
\tabletypesize{\scriptsize}
\tablecolumns{8}
\tablewidth{0pt}

\tablecaption{SMC Radial Velocity Studies \label{tab:rvsamples}}
\tablehead{
    \colhead{Reference} & \colhead{$v_{r,helio}$} & \colhead{$\sigma$} & 
        \colhead{sample description} \\
    \colhead{} & \colhead{[km~s$^{-1}$]} & \colhead{[km~s$^{-1}$]} &
        \colhead{} 
}

\startdata
\cite{dop85} & $146\pm4$ & $25\pm3$ &   44 planetary nebulae                   \\
\cite{sun86} & $123$     & $24.2$   &   12 red giant stars near NGC~121        \\
\cite{mau87} & $162\pm1$ & $18$     &  255 supergiants and main sequence stars \\
\cite{mat88} & $149\pm3$ & $22\pm3$ &   61 cepheid stars                       \\
\cite{har89} & $148\pm4$ & $27\pm2$ &  131 central carbon stars                \\
\cite{hat93} & $151\pm6$ & $33\pm4$ &   29 red clump stars in outer NE quadrant\\
\cite{ss97}  & $155\pm1$ & $25\pm1$ &  501 expanding HI shells         \\
\cite{hat97} & $149\pm3$ & $26\pm2$ &   72 outer carbon stars                  \\
this work    & $145.6\pm0.6$ & $27.6\pm0.5$ & 2046 red giant stars         \\
\enddata
\end{deluxetable}

\clearpage

\end{document}